# A low cost non-wearable gaze detection system based on infrared image processing

Ehsan Arbabi [1], Mohammad Shabani [2] and Ali Yarigholi [3]

School of Electrical and Computer Engineering, College of Engineering, University of Tehran, Iran

[1] *earbabi@ut.ac.ir,* [2] *mshabani69@alumni.ut.ac.ir,* [3] *ali.yarigholi@alumni.ut.ac.ir*

[1,2,3] All authors contributed equally to this work.

**Abstract:** Human eye gaze detection plays an important role in various fields, including human-computer interaction, virtual reality and cognitive science. Although different relatively accurate systems of eye tracking and gaze detection exist, they are usually either too expensive to be bought for low cost applications or too complex to be implemented easily. In this article, we propose a non-wearable system for eye tracking and gaze detection with low complexity and cost. The proposed system provides a medium accuracy which makes it suitable for general applications in which low cost and easy implementation is more important than achieving very precise gaze detection. The proposed method includes pupil and marker detection using infrared image processing, and gaze evaluation using an interpolation-based strategy. The interpolation-based strategy exploits the positions of the detected pupils and markers in a target captured image and also in some previously captured training images for estimating the position of a point that the user is gazing at. The proposed system has been evaluated by three users in two different lighting conditions. The experimental results show that the accuracy of this low cost system can be between 90% and 100% for finding major gazing directions.

## 1   Introduction

Using eye tracker is common in different research and applications related to human-computer interaction, intelligent systems, virtual reality, psychology and cognitive science. For instance, it is suggested to use eye tracker as an input device for computer control [1-3]. In this case a user can run applications software or manage peripheral devices by looking at related menu options displayed on a screen [3]. Slambekova et al. (2012) presented a framework for interaction within a 3D virtual world, which enables the use of eye gaze and hand gestures [4]. Tanriverdi and Jacob (2000) also proposed an eye movement-based interaction technique in a virtual environment. In addition, they compared their proposed technique to more conventional 3-D pointing and concluded that the eye movement-based interaction can be faster than pointing, especially for distant objects [5]. Bee et al. (2010) introduced a framework using various set of sensors, including eye tracker, in order to analyze the users' state which then can influence the feelings of a virtual characters and their actions in a story [6]. In another work, the same authors presented a system for interpreting users' attentive state through eye gaze, while interacting with a virtual character [7].

Mele and Federici (2012) provided an overview of the main psychological applications of eye-tracking. In their article, experimental tasks such as visual search, reading, and scene viewing have been considered [8]. Sanchez et al. (2013) used eye tracking for depressed participants in order to examine difficulties related to disengaging attention from emotional material [9]. Pavlov et al. (2014) evaluated the influence of meditation on attentional bias towards neutral and emotional facial expressions by eye tracking [10]. Reichle et al. (2010) examined eye movements during mindless reading, which occurs when the eyes continue moving across the page while the person thinks about something unrelated to the text [11]. Di Giorgio et al. (2012) recorded eye movements in infants and adults in order to investigate the ability for detecting and preferring a face when embedded in complex visual displays [12].

Finally, different systems, utilizing eye tracking, have been applied for evaluating driver fatigue and alertness [13-17]. For example, Ji and Yang (2002) applied eye, gaze, and face pose tracking techniques by using an infrared camera in order to monitor driver vigilance [17].

For eye tracking, different techniques usually based on measuring bioelectrical potential, optical reflection or mechanical displacement, have been developed. For example by recording and processing Electrooculogram (EOG) signals, the orientation of users' eyeballs can be found [18, 19]. Also, by attaching trackable contact lenses on eyes (e.g. using scleral search coil technique), it is possible to measure eye movements [20, 21]. Finally by using optical methods, such as capturing either visual or infrared images of eyes, eye movements can be measured [17, 22]. In all these methods after estimating eyes orientation, by knowing the head position, it is possible to find the regions that a user is gazing at.

The main differences between eye trackers, regardless of their technologies, can be summarized in four issues: 1- accuracy, 2- complexity (and price), 3- ability of head tracking, and 4- comfortability. In some medical applications, the accuracy of pupil/gaze detection can be considered as the most important issue. Therefore using complex and expensive instruments for gaining the desired accuracy is reasonable. However, there are many other applications in which just finding the approximated gazing regions is enough. For example in some applications related to cognitive science or virtual reality, it may be important to know whether a user is looking at a special area or not. Thus, in these cases, using highly accurate and consequently complex and expensive systems is unnecessary.

After estimating eye orientation, the gazing region can be found based on head position. If users are allowed to move their head during eye tracking, performing other measurements for head tracking is needed.

Therefore, all eye tracking systems which only provide eye orientation as their output, are solely useful for detecting gazing regions in fixed-head situations.

Many eye trackers need to be mounted on head. For example, the systems based on EOG signal processing or contact lens tracking are in this category [18-21]. In these cases, the attaching parts (e.g. lenses or recording electrodes) need to be placed carefully on their designated positions, which consequently can make users feel less comfortable during utilizing the devices, comparing to normal situation. Although some optical eye tracking devices are also mounted on head [23], non-wearable eye tracking devices exists too [17, 24-26]. Thus, it seems that if users' convenience is an important factor in the target applications (e.g. in daily/long term usage applications or some psychological/cognitive science research), using optical based eye tracking methods should be considered.

There are many optical eye tracking devices which can provide measurements with high accuracy; however, their implementation may be either relatively complex or expensive for regular users and applications. For example, many optical methods are based on pupil–corneal reflection technique [17, 24, 26-28] and may require using relatively high quality capturing system and also synchronizing camera shutter with lighting sources, in order to produce and detect corneal glints. Other technical conditions, such as using more than one camera, special type of camera (e.g. Kinect or moving camera), different external lenses, and ultrasonic distance meter may also reduce the affordability and/or ease of implementation of such devices for many users [22, 24-26, 29].

The work presented in this paper is originally based on a research done in 2013 in the School of Electrical and Computer Engineering at University of Tehran. We propose a new optical based eye tracking system with low complexity and cost, high comfortability, ability of head tracking and medium accuracy. Therefore, the novelty of this work, in opposite to many researches, is not the accuracy. In other words, the proposed system may not be suitable for highly accurate applications (such as some medical cases). But due to its low complexity and cost, it is easy to be implemented and used in different research and applications with medium range of accuracy. Also, the main system is not mounted on head or any part of body, which makes it more convenient for users. Of course, in order to add head tracking ability, three small pieces of retro-reflective adhesive paper are needed to be placed on users' face, which normally should not affect their convenience.

The proposed method includes two main processing steps: 1- pupil detection and 2-gaze evaluation. In the first step, by using infrared images, captured by a simply modified regular visible webcam, and applying image processing techniques, positions of retro-reflective adhesive papers (markers) and eye pupils are detected. In the second step, by applying a training phase on some pre-captured images, the eye gazing region for a current captured image is estimated.

The remaining of the article is as follows. Different steps of the proposed method are explained in section 2. The details about the experimental evaluation of the method on the data captured from three subjects in two different lighting conditions, is covered in section 3. In section 4, the obtained experimental results are reported and discussed. Finally, the conclusion of the paper is drawn in section 5.

## 2 Methods

The proposed methods consist of two main steps, i.e. 1- image processing step for pupil and marker detection, and 2- data analysis step for gaze detection.

### 2.1 Pupil detection

In this step, the facial image (which can be either a solo-image or a video frame) is captured by using infrared camera. Since we want to capture infrared images, it is important to have infrared lighting sources. By positioning the infrared light emitters around the camera lens, retro-reflective objects can be distinguished from non-retro-reflective passive objects, due to their higher intensity in the captured images. Thus, because of human eyes retro-reflective characteristic, the pupils are usually bright in the captured infrared images [17]. In our method, in addition to using infrared camera with infrared light emitters around its lens, three small pieces of retro-reflective adhesive papers are placed on users' face. The markers are basically used for head tracking during gaze detection step; however, we can take advantage of their presence for pupil detection too. Marker #1 (#2) is placed just above the right (left) corner of the right (left) eyebrow. Marker #3 is placed just below the area between two eyes and above the nose. By applying this setup, we expect to see pupils and markers brightly in the captured images (see Fig. 1).

By avoiding any active light source in the viewing range of camera, the retro-reflective markers show higher intensity in the captured images, comparing to other elements, including the pupils (which are weaker retro-reflective elements). Therefore, the markers can be easily detected by using a suitable threshold in gray-scaled images. For this reason, first histogram equalization is applied on the gray-scaled image. Then, N numbers of pixels with the highest intensity are found in the histogram-equalized image. The value of N is chosen in a way to ensure that the selected pixels include the markers. This value can be either found experimentally or considered equal to the maximum possible number of pixels for the markers (it can be done by evaluating a sample image prior to the main processing). The lowest intensity of these N chosen pixels is selected as a threshold for producing a binary image. In the output binary image, the markers are white and thus their locations are found and stored.

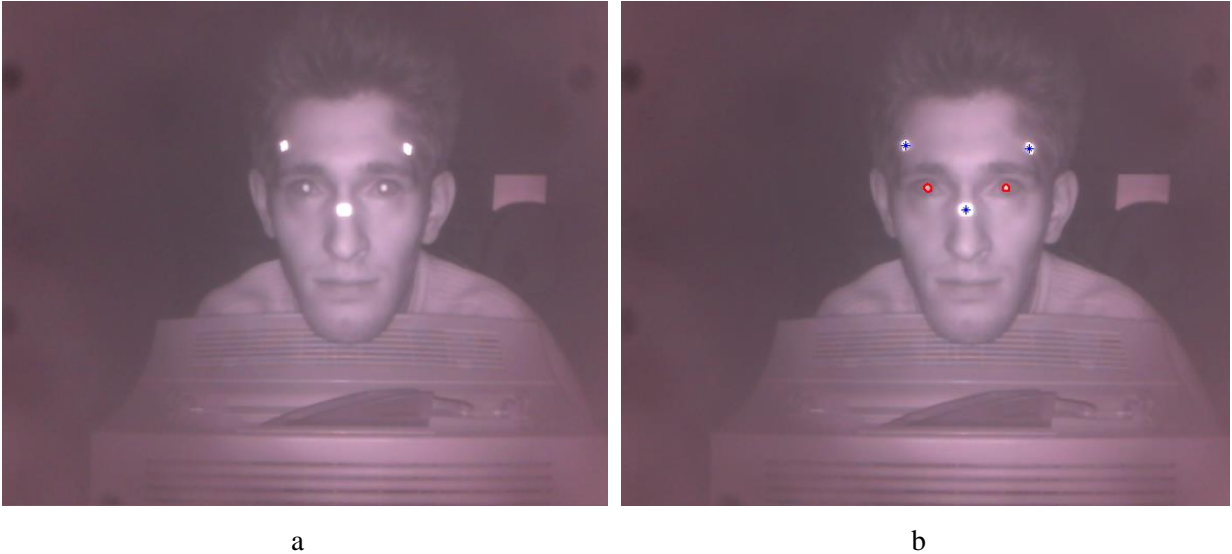

a　　　　　　　　　　　　　　　　　　　　　　　　b

Fig. 1. A sample infrared image, in which the pupils and markers are bright due to retro-reflection: a) the raw input image; b) the image after markers and pupils detection.

By knowing the positions of the markers, our regions of interests for finding the pupils can be restricted. For finding the right (or left) pupil, a rectangular region is extracted from the image by considering maker #1 (or #2) and marker #3 as diagonal corners. The weak retro-reflective character of pupil and also the reflective character of cornea make pupil detection more challenging than finding the markers. Due to the position of central marker (#3) on nose, removing its pixels from the regions of interests may crop eye images. Therefore, in our method, we make sure that the regions of interests exclude only the pixels related to the right and left markers (Fig. 2.a). After histogram equalization of the extracted rectangular region (Fig. 2.b), a threshold value, larger than mean intensity of the region, is chosen. In our method, we choose this value as the weighted average intensity of the extracted region, in which the intensity of the pixels with higher intensity than the mean intensity is multiplied by two (chosen experimentally). Applying this threshold on the extracted region provides a binary image in which some areas including pupil are white (Fig. 2.c or Fig. 2.f). By using morphological closing operator, with element size equal to 10% of approximated pupil size, we remove all tiny white areas resulted from artifacts or possible cornea reflection (Fig. 2.d or Fig. 2.g). Since the pupil is expected to be completely inside the extracted rectangular region, any white area connecting to the borders (such as skin) are also removed. Finally, the areas with eccentricity [30] less than 0.9 is detected as pupil (Fig. 2.e or Fig. 2.h). The locations of pupils and markers are represented in a mathematical Cartesian coordinate system, in which X and Y axes are in horizontal and vertical directions, respectively. The location of each pupil or marker is considered as the location of the middle pixel in each detected region. These locations are represented as $(x_{M,r}, y_{M,r})$, $(x_{M,m}, y_{M,m})$, $(x_{M,l}, y_{M,l})$, $(x_{P,r}, y_{P,r})$ and $(x_{P,l}, y_{P,l})$, for x and y coordinates of right marker, middle marker, left marker, right pupil and left pupil, respectively. By right/left we mean the right/left side of the captured image, so that $x_{(.),r} \geq x_{(.),l}$. In order to decrease wrong pupil detection rate (false positive) caused by vertically far mis-detected pupils, the vertical positions of two detected pupils are compared to each other applying the condition represented in Eq. (1);

$$(y_{P,r} - y_{P,l})^2 \leq 0.25 \times \left| \left(y_{M,r} - y_{M,l}\right)\left(y_{M,m} - \left(\frac{y_{M,r} + y_{M,l}}{2}\right)\right) \right|. \tag{1}$$

It should be mentioned that in case of detecting more than one pupil in the extracted region (Fig. 2.e), the algorithm restarts by using higher value of threshold. The output images in different steps of pupil detection procedure are demonstrated in Fig. 2.

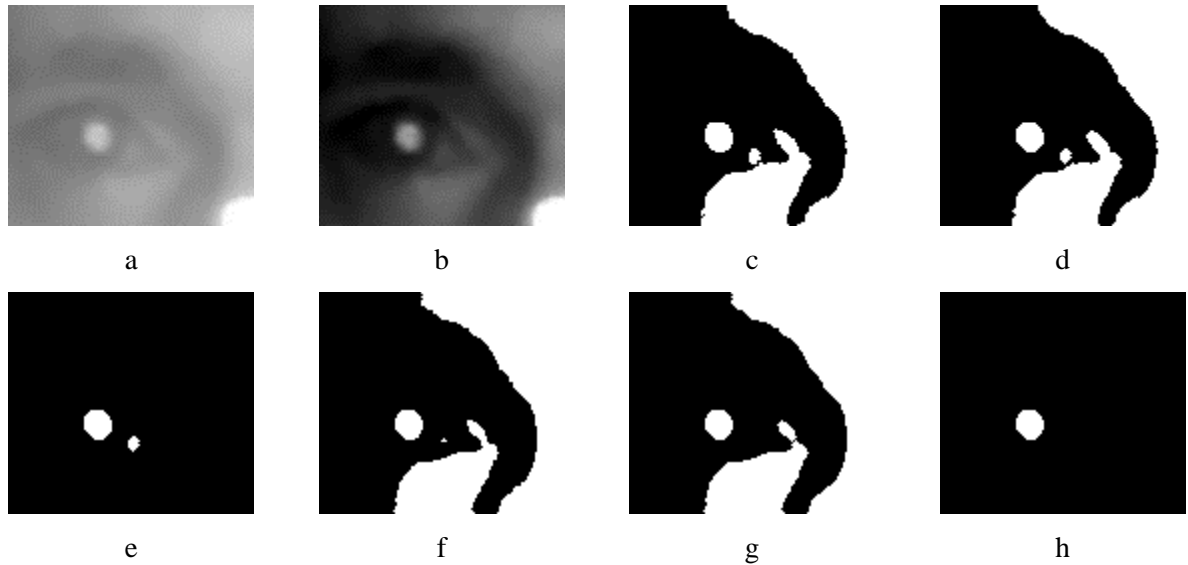

Fig. 2. Pupil detection procedure: a) gray-scaled extracted region of interests; b) after histogram equalization; c) the first binary image prepared by using weighted average intensity as the threshold value; d) the first binary image after applying the closing morphological operator; e) the detected pupil (two pupils: wrong answer); f) the second binary image using a new threshold value (higher than the previous one); g) the second binary image after applying the closing morphological operator; h) the detected pupil (correct answer).

## 2.2 Gaze estimation

After detecting markers and pupils, it is time to find the gazing point. Our proposed method for gaze detection is based on system training. We first provide some known training data containing pupils and markers locations on a captured image along with the location of the point that the user was gazing at during image capturing. For this reason, before starting eye tracking, we ask the user to gaze at some certain and known points. In our study, the user should gaze at four corners of the target screen (i.e. the screen that the user is going to watch/read during eye tracking). The user is also asked to slowly move and rotate her/his head in different arbitrary directions and orientations, while maintaining gazing at one of the known points. The infrared images of the user's face are captured continuously during this training period and it is repeated for all the known points (i.e. four corners of the target screen). Thus, we have collection of images (frames) with known gazing points for each user, separately. For each training image, pupils and markers locations on the captured image are estimated by using the proposed image processing method (explained in section 2.1). We consider the training gazing points as the up-left, up-right, down-left and down-right corners of the target screen and represent them with index values of 1, 2, 3 and 4, respectively. The total numbers of infrared images, captured during gazing at these corners (training period) are represented by $N_1$, $N_2$, $N_3$ and $N_4$, respectively. Finally, each training vector is represented as Eq. (2);

$$\text{Training Vectors (c, i)} = [x_{M,r}^{c,i}, y_{M,r}^{c,i}, x_{M,m}^{c,i}, y_{M,m}^{c,i}, x_{M,l}^{c,i}, y_{M,l}^{c,i}, x_{P,r}^{c,i}, y_{P,r}^{c,i}, x_{P,l}^{c,i}, y_{p,l}^{c,i}]$$
$$c = 1, 2, 3, 4 \quad \& \quad i = 1, 2, \ldots, N_c; \quad (2)$$

in which c corresponds to the corner index of the target screen that the user was gazing at, when the image was captured.

During eye tracking process, infrared images are captured while the user changes her/his gazing direction during watching/reading the target screen. For each input captured image, the pupils and markers locations are again estimated by applying the proposed image processing algorithm. Our aim is to use these locations and also the training vectors in order to estimate the location of the gazing point on the target screen, corresponding to the input captured image. For clarity, we represent all x and y coordinates of the pupils and the markers in the input captured images by x′ and y′. Our first step for finding the gazing point, is to find the closest head orientation in the training set to the head orientation of the input image. For this reason, we compare the locations of the markers in our training set and input image. This comparison is done for the training sets corresponding to each of four corners separately. Therefore, at the end we have four training vectors corresponding to gazing at four different corners of the target screen, in which the head orientations are the closest to the head orientation of the input image. Normally, for a complete training set, using Euclidian distance between the markers in the training data and their corresponding markers in the input image should be sufficient for finding the closest head orientation to the input image. However, due to possible unavailability of complete or semi-complete training set, experimentally we found out that using only Euclidian distance may not be helpful. In our method, we have used geometrical congruency between the triangles composed by the markers as a measure for finding the closest head orientation. This measure is defined in Eq. (3), in which A (A′), B (B′) and C (C′) are the length of the edges in the triangle composed by three markers (as its vertices), in a training image (the input image);

$$M = 3 - (A/A' + B/B' + C/C'). \tag{3}$$

When comparing two training vectors, the one with smaller absolute value for M is chosen as the closest one to the input image. Since our concentration is on avoiding making the system difficult for implementation or usage by common users, preparing complete or semi-complete database is ignored. However, for complete or semi-complete database, we recommend using Euclidian distance with/out geometrical congruency for measuring similarity between head orientations.

After finding the closest training vectors (one vector for each corner), we need to transform all the coordinate values of the found training vectors (pupils and markers) in order to reduce the effect of head translation. For this transformation, different geometric strategies have been tested. However, experimentally we found out that simple translation based on the location of the middle marker provides the best matching between the input image and four closest training vectors. In fact, it seems that due to the position of middle marker, which is almost in the central part of face, the natural head translations of human can be better predicted by evaluating the translation of the middle marker, comparing to the other markers (or even their combinations). Thus, we transform all x and y coordinates of the found four training vectors in a way that the locations of the middle markers in all five images (four training and one input) become the same. We denote the coordinates of the right and left pupils in the images corresponding to gazing at corner c (= 1, 2, 3, 4) and the input image by $\{(x^c_{P,r}, y^c_{P,r})\ \&\ (x^c_{P,l}, y^c_{P,l})\}$ and $\{(x'_{P,r}, y'_{P,r})\ \&\ (x'_{P,l}, y'_{p,l})\}$, respectively.

Now we have four modified (translated) training vectors, corresponding to gazing at four corners that their head orientation is close to the head orientation of the input image. By knowing the pupils locations and also gazing points (corners) in these four modified training vectors, we estimate the gazing point of the input image based on its pupils locations. For simplicity, the coordinates of the gazing point on the target screen for the training images (gazing at corner c) and the input image (to be found) are denoted by $(x^c_G, y^c_G)$ and $(x'_G, y'_G)$, respectively. The coordinate system used for representing the gazing points on the target screen is also a mathematical Cartesian coordinate system, in which X and Y axes are in horizontal and vertical directions, respectively. In our method, $(x'_G, y'_G)$ can be found by using coordinates of either right pupils or left pupils of all images. For simplicity, we first explain our proposed strategy using coordinates of the right pupils in all images. Applying linear interpolation, we can estimate $x'_G$ by using either Eq. (4) or Eq. (5);

$$x'_G = \alpha \left(x^2_G - x^1_G\right) + x^1_G, \tag{4}$$

$$x'_G = \beta \left(x_G^4 - x_G^3\right) + x_G^3, \tag{5}$$

in which α and β are estimated by using pupils locations corresponding to the related corners, based on Eq. (6) and Eq. (7);

$$\alpha = (x'_{P,r} - x_{P,r}^1)/(x_{P,r}^2 - x_{P,r}^1), \tag{6}$$

$$\beta = (x'_{P,r} - x_{P,r}^3)/(x_{P,r}^4 - x_{P,r}^3). \tag{7}$$

Normally, the target screen is square. In this case, $y_G^1 = y_G^2$, $y_G^3 = y_G^4$, $x_G^1 = x_G^3$ and $x_G^2 = x_G^4$. Thus, Eq. (4) and Eq. (5) become almost similar. However, experimentally, α and β are not the same. It means, Eq. (4) and Eq. (5) provide different estimations. Therefore, it seems rational to take a weighted average of these two estimations, similar to Eq. (8);

$$x'_G = W\left(\alpha\left(x_G^2 - x_G^1\right) + x_G^1\right) + (1-W)\left(\beta\left(x_G^4 - x_G^3\right) + x_G^3\right), \quad 0 \leq W \leq 1 \tag{8}$$

in which W is the weight which should be chosen based on the approximate gazing point in the input image. In fact, using Eq. (4) or Eq. (5) are more suitable when the user is gazing at upper half or lower half of the target screen, respectively. Therefore, W should be inversely proportional to the distance between the y coordinate of the gazing point on the target screen ($y'_G$) and the y coordinate of the upper edge of the target screen ($y_G^1$ or $y_G^2$). Since $y'_G$ is unknown, instead of using $y'_G$ and $y_G^c$, we use $y'_{P,r}$ and $y_{P,r}^c$ for estimating W:

$$W = \left(y'_{P,r} - \frac{y_{P,r}^3 + y_{P,r}^4}{2}\right) \Big/ \left(\frac{y_{P,r}^1 + y_{P,r}^2}{2} - \frac{y_{P,r}^3 + y_{P,r}^4}{2}\right). \tag{9}$$

By using a similar procedure, $y'_G$ can also be estimated based on Eq. (10) to Eq. (13).

$$y'_G = W'\left(\gamma\left(y_G^1 - y_G^3\right) + y_G^1\right) + (1-W')\left(\delta\left(y_G^2 - y_G^4\right) + y_G^4\right), \tag{10}$$

$$\gamma = (y'_{P,r} - y_{P,r}^3)/(y_{P,r}^1 - y_{P,r}^3), \tag{11}$$

$$\delta = (y'_{P,r} - y_{P,r}^4)/(y_{P,r}^2 - y_{P,r}^4), \tag{12}$$

$$W' = \left(x'_{P,r} - \frac{x_{P,r}^1 + x_{P,r}^3}{2}\right) \Big/ \left(\frac{x_{P,r}^2 + x_{P,r}^4}{2} - \frac{x_{P,r}^1 + x_{P,r}^3}{2}\right). \tag{13}$$

By now, we could estimate ($x'_G$, $y'_G$) by only using the coordinates of the right pupils. Similarly, we can estimate ($x'_G$, $y'_G$), using the coordinates of the left pupils. In this case instead of using $x_{P,r}^c$ and $x'_{P,r}$, $x_{P,l}^c$ and $x'_{P,l}$ are used respectively for calculating α, β, γ, δ, W and W′ (c = 1, 2, 3, 4). One benefit of this situation is that if a pupil is not detected properly in the input image, we can still estimate ($x'_G$, $y'_G$), using the coordinates of another detected pupil. When both pupils are detected, ideally we expect that the equations related to the right pupils and the left pupils result in the same answer for ($x'_G$, $y'_G$). However, due to different factors such as possibility of poor pupil detection and neglecting eyeball radii, the resulted ($x'_G$, $y'_G$) by the right and the left pupils are not necessarily the same. Thus, in our method we consider the average of the right and the left pupils for estimating the gazing point.

## 3   Experiments

### 3.1   System setup

In order to evaluate the proposed methods, first, a webcam (Logitech – C250) with snapshots at up to 1.3-megapixels was chosen for image capturing. Its IR filter was replaced with a visible light filter, in order to transform the webcam into an infrared camera. A small piece of darkened analog photographic film was used as the visible light filter. It should be added that other materials such as magnetic material inside Floppy disks can also be used as visible light filter [31]. For exploiting the retro-reflective effects of the pupils and the markers, six IR LEDs (precise wavelength specification is not required) were placed around the camera lens. These IR LEDs were connected to a power supply to be constantly on, during eye tracking (see Fig. 3). Finally, the prepared infrared camera was positioned about 65 cm far from the users' face.

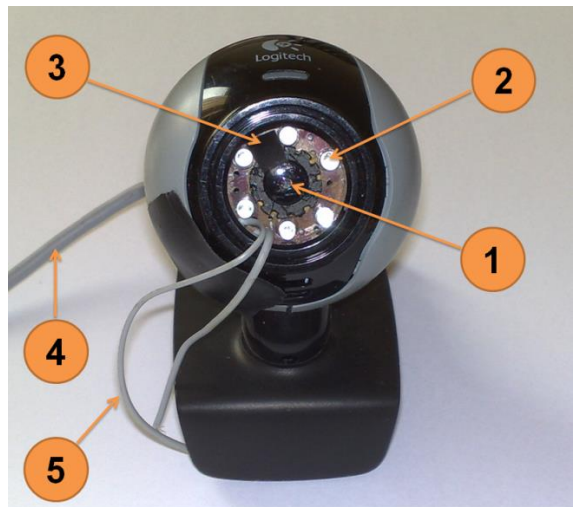

Fig. 3. The infrared camera prepared for our experiments by modifying a common webcam: 1) lens, 2) implemented IR-LED, 3) visible light filter (should be moved to cover the lens), 4) data cable to be connected to a computer, 5) power supply cables for IR-LEDs.

E Arbabi, M Shabani & A Yarigholi; *A low cost non-wearable gaze detection system based on infrared image processing*; Available online: 2017.

For the target screen, a screen with visible rectangular grids (5 by 5) was chosen. The target screen was consisted of twenty-five 12 cm * 12 cm square cells, similar to Fig. 4.

| #1 | #2 | #3 | #4 | #5 |
|----|----|----|----|----|
| #6 | #7 | #8 | #9 | #10 |
| #11 | #12 | #13 | #14 | #15 |
| #16 | #17 | #18 | #19 | #20 |
| #21 | #22 | #23 | #24 | #25 |

Fig. 4. A 5 by 5 target screen and the labels of its cells.

We tested the proposed algorithms on three male subjects with normal sight in two different lighting conditions. In the first lighting condition, the room, where the experiments were carried on, was lighted with the minimum required light for seeing the target screen (low light). In the second lighting condition, the room was lighted normally, regardless of the experiment (normal light). Each user was sat in a way that he could see the middle cell (#13) of the target screen when looking forward (without any head rotation). The distance between the users' eyes and the target screen was set to be about 65 cm. By this setup, the vertical and horizontal angular range of vision (field of view) for each user was considered as about 50° in its maximum situation (when a user faces and gazes forward).

## 3.2 Data collection

Three retro-reflective markers were placed on each user's face according to the previous explanations (see Fig. 5). Each user was asked to gaze at the middle point of every cell only by moving his eyes, while keeping his head almost steady in its position. This was done for six different head positions, and two different lighting conditions. In these six head positions, the users were almost facing (not gazing) at cells #5, #6, #13, #15, #17 and #23, which cover the main head orientations in a semi-random manner. Finally, we captured totally 900 infrared facial images (25 gazing point * 6 head positions * 2 lighting conditions * 3 users). A sample captured infrared image is shown in Fig. 1.a. Due to some problems, such as misplacements of the markers for the third user, we had to remove 28 images from his 150 captured images in normal lighting condition.

The proposed algorithms were evaluated offline by using the captured images. In fact, first the pupils and the markers positions have been found for all images. Then, by using the images related to gazing at the corners of the target screen (cells #1, #5, #21 and #25), we used the proposed interpolation based algorithm to estimate the gazing position on the target screen for each image.

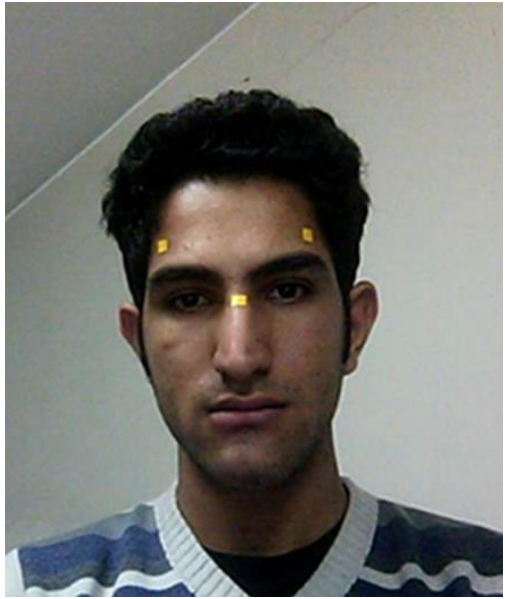

Fig. 5.  Three retro-reflective adhesive papers (in yellow) are placed on a user's face for tracking head orientations.

## 4   Results and Discussion

The estimated gazing point in each captured image has been compared to the real position of the gazing point and the error in both X and Y directions have been found (we call them Δx and Δy respectively). Based on Δx and Δy, we can evaluate the total system accuracy for different resolutions. If the target screen has the resolution of N by N (5 by 5 in our case), Δx and Δy should satisfy the conditions mentioned in Eq. (14), in order to consider the estimated gazing point correct;

$$|\Delta x| < L_x/(2*N) \quad \& \quad |\Delta y| < L_y/(2*N), \tag{14}$$

in which $L_x$ and $L_y$ are the length of the target screen in X and Y directions, respectively. Finally, the accuracy of the system is represented by dividing the number of images in which the gazing cell is found correctly to the total number of images. Although in our experiments we used a target screen with the resolution of 5 by 5, we still can use other values for N in order to approximately evaluate the accuracy of the system in other resolutions. In our study, we used the values between 2 and 10 for N. The accuracies of the proposed system in low lighting condition and normal lighting condition are shown in Table 1 and Table 2, respectively. In these tables AVG and STD are average and standard deviation of the accuracy for all three users, respectively.

Table 1. The accuracy (%) for different resolutions of the target screen in low lighting condition.

| Resolutions | User 1 | User 2 | User 3 | AVG ± STD |
|---|---|---|---|---|
| 2 by 2 | 100.0 | 100.0 | 100.0 | 100.0 ± 0.0 |
| 3 by 3 | 100.0 | 100.0 | 100.0 | 100.0 ± 0.0 |
| 4 by 4 | 98.7 | 100.0 | 98.7 | 99.1 ± 0.8 |
| 5 by 5 | 97.3 | 98.0 | 98.0 | 97.8 ± 0.4 |
| 6 by 6 | 93.3 | 92.0 | 96.7 | 94.0 ± 2.4 |
| 7 by 7 | 89.3 | 87.3 | 94.0 | 90.2 ± 3.4 |
| 8 by 8 | 85.3 | 82.0 | 85.3 | 84.2 ± 1.9 |
| 9 by 9 | 76.0 | 80.0 | 78.7 | 78.2 ± 2.0 |
| 10 by 10 | 70.0 | 74.7 | 74.0 | 72.9 ± 2.5 |

Table 2. The accuracy (%) for different resolutions of the target screen in normal lighting condition.

| Resolutions | User 1 | User 2 | User 3 | AVG ± STD | AVG ± STD (excluding user 3) |
|---|---|---|---|---|---|
| 2 by 2 | 97.3 | 97.3 | 90.2 | 94.9 ± 4.1 | 97.3 ± 0.0 |
| 3 by 3 | 96.7 | 94.7 | 83.6 | 91.7 ± 7.0 | 95.7 ± 1.4 |
| 4 by 4 | 96.0 | 92.7 | 82.8 | 90.5 ± 6.9 | 94.3 ± 2.4 |
| 5 by 5 | 94.7 | 91.3 | 79.5 | 88.5 ± 8.0 | 93.0 ± 2.4 |
| 6 by 6 | 91.3 | 90.7 | 72.1 | 84.7 ± 10.9 | 91.0 ± 0.5 |
| 7 by 7 | 86.0 | 88.7 | 66.4 | 80.4 ± 12.2 | 87.3 ± 1.9 |
| 8 by 8 | 82.0 | 82.7 | 61.5 | 75.4 ± 12.1 | 82.3 ± 0.5 |
| 9 by 9 | 78.0 | 76.0 | 54.1 | 69.4 ± 13.3 | 77.0 ± 1.4 |
| 10 by 10 | 75.3 | 72.7 | 48.4 | 65.5 ± 14.9 | 74.0 ± 1.9 |

As it can be seen, the accuracy of gaze detection for the third user in normal light condition is considerably less than the other users. In fact, the images related to this user usually had poor quality and the markers were not placed properly. It is why that we have removed 28 images from his dataset. If we remove his dataset completely from our evaluation, the average and the standard deviation of the accuracy increase and decrease respectively (see the last column of Table 2). In order to compare two lighting conditions with each other, the average accuracies of two lighting conditions are illustrated in Fig. 6.

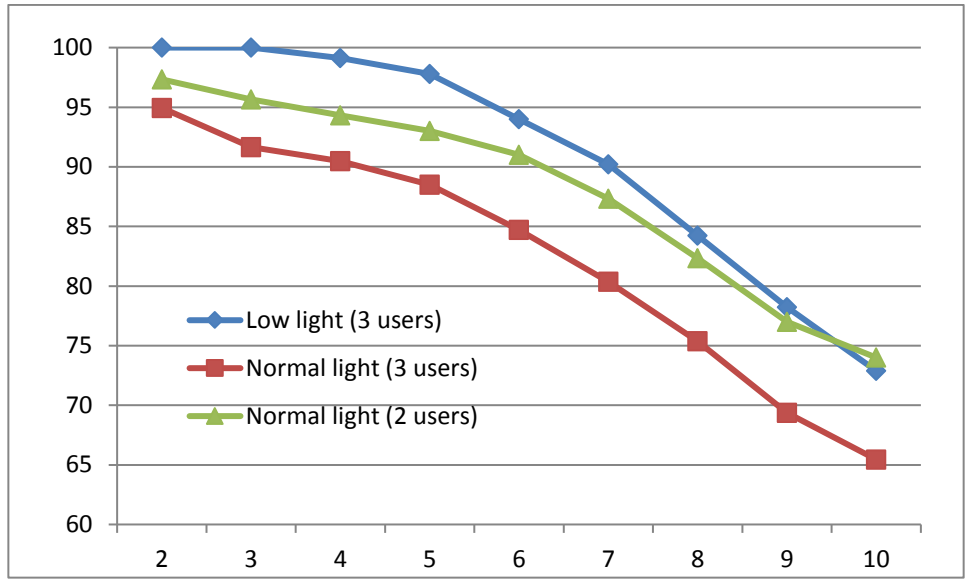

Fig. 6. Accuracy of the proposed gaze detection system for different resolutions of a target screen. Vertical and horizontal axes demonstrate the accuracy (%) and the resolution (N), respectively.

Based on the results it can be seen that in cases when only knowing general gazing point is important the proposed algorithm works well. For example, if we only want to know whether a user is mainly gazing up, down, center, right or left, we maximally need the resolution of 3 by 3. As it can be seen in Table 1 and Table 2, the accuracy for this resolution is between 90% and 100% depending on lighting conditions. Even, for finer resolutions the accuracy can still be more than 90%. More specifically, in our experiments the average accuracy was more than 90% in 7 by 7 and 4 by 4 (6 by 6) resolutions, for low and normal (excluding the 3[rd] user) lighting conditions, respectively. Based on Fig. 6, it can be observed that the accuracy in two lighting conditions approaches each other when increasing the resolution (excluding the 3[rd] user for normal lighting). It means that for higher resolutions (in opposition to lower resolutions), the provided lighting conditions are not playing the important role in gaze detection.

It should be noticed that in our setup the maximum field of view is about 50°, which provides a convenient situation for the users. Thus, in case we increase the maximum field of view, it can be expected to have a much better accuracy in finer resolutions. Also, as it can be seen in Fig. 1, due to the camera characteristics and position (simple and non-wearable), the region of interests covers a small portion of the captured image. It is clear that if one uses a camera with higher resolution and in a closer position to the face, the accuracy can be improved for much finer resolutions.

In the applications that the gazing point should be detected continuously, the proposed algorithm can be combined with a predicting algorithm (such as Kalman filter) in order to improve the accuracy. In fact, in continuous gaze detection, the current gazing point can also be predicted by the other gazing points found in the previous frames.

Concerning the processing speed, due to the algorithm simplicity, the proposed method can be used for semi-real-time applications if needed. In our experimental test, the method could estimate gazing point for a captured image in less than 200 ms (mainly devoted to markers and pupils detection step), using MATLAB 7.12.0 (32-bit) and an Intel Core i5-2300 CPU @ 2.8 GHz. It should be pointed out that using other programming software such as C++ can improve the processing speed.

## 5  Conclusion

In this article, we have proposed a new method for gaze detection by using infrared images. The main contribution is using both low cost and easy-to-implement hardware and algorithms for achieving a proper accuracy in general applications. The experimental results have shown that the accuracy of the system can be close to 100% for finding the main gazing directions (e.g. 3 by 3 resolution), even when considering the maximum field of view as 50°. In addition to simplicity and low cost of our non-wearable system, our eye tracker is also able to consider head's movement for gaze detection.

Many research works have tried to improve the accuracy of eye tracking systems. However, their proposed system may be head mounted, complex (expensive) and/or without head tracking ability. Of course, by reviewing the literatures we could find eye tracking systems (such as the system proposed in [17]), which are also able to perform head tracking without any need to be mounted on head. These systems can be certainly well suited for different applications; however, they may need high quality of images and consequently more advanced camera (comparing to our simple and low cost system) in order to work properly. Also, due to the oscillatory behavior of some of these systems (IR-LEDs need to be turned on and off continuously in high rate), they are more complex and costly comparing to our system.

Finally, in terms of accuracy we still consider other technologies as superior. But, regarding other main issues (especially comfortability, complexity and cost), we think the proposed system is an appropriate choice for the applications with medium range of accuracy. In other words, the proposed strategy can be helpful for making eye-tracking easily available for the research and applications, in which gaze detection with very high accuracy (using expensive and/or complicated systems) is useless.

As future works, evaluating other interpolation strategies can be considered for gazing point estimation. In fact, an important issue in our method, which can affect the accuracy, is the interpolation strategy that we used for gaze detection. Using linear interpolation, which may result in more stability comparing to non-linear interpolations, helped us to show that the proposed method which is fundamentally based on exploiting training data, works properly for gaze detection in general cases. Thus, evaluating other types of non-linear interpolations can be considered as a next step for possible improvement of the system.

## Acknowledgments

We would like to thank Mr. Ramtin Ardeshiri, Mr. Mohammad Reza Mohagheghi-Nejad and all the students at University of Tehran who assisted us in this project in different ways, including participating in data collection process.

This work has been funded by Iranian National Science Foundation (INSF).